\documentclass[lettersize,journal]{IEEEtran}
\usepackage{amsmath,amsfonts,amssymb}
\usepackage{algorithmic}
\usepackage{algorithm}
\usepackage{array}
\usepackage[caption=false,font=normalsize,labelfont=sf,textfont=sf]{subfig}
\usepackage{textcomp}
\usepackage{stfloats}
\usepackage{url}
\usepackage{verbatim}
\usepackage{graphicx}
\usepackage{cite}
\usepackage{cleveref}
\usepackage{tabularx}
\usepackage{makecell}
\usepackage{multirow}
\usepackage{booktabs}
\usepackage{enumitem}
\usepackage{balance}
\usepackage{soul}
\usepackage[table]{xcolor}

\usepackage{amsfonts,bm}

\def\eqref#1{equation~\ref{#1}}

\def\1{\bm{1}}

\def\vk{{\bm{k}}}

\def\vq{{\bm{q}}}

\def\vz{{\bm{z}}}

\def\mA{{\bm{A}}}
\def\mB{{\bm{B}}}
\def\mC{{\bm{C}}}

\def\mE{{\bm{E}}}

\def\mW{{\bm{W}}}

\def\mZ{{\bm{Z}}}
\def\mAlpha{{\bm{\alpha}}}

\DeclareMathAlphabet{\mathsfit}{\encodingdefault}{\sfdefault}{m}{sl}
\SetMathAlphabet{\mathsfit}{bold}{\encodingdefault}{\sfdefault}{bx}{n}

\def\gL{{\mathcal{L}}}

\newcommand{\R}{\mathbb{R}}

\newcommand{\softmax}{\mathrm{softmax}}

\newcommand{\method}{\textsc{LDGA}}
\newcommand{\dataset}[1]{\texttt{#1}}

\newtheorem{definition}{Definition}

\begin{document}

\title{Layered Division and Global Allocation for Community Detection in Multilayer Network}

\author{Fanghao Hu, Zhi Cai, and Bang Wang$^{*}$
\thanks{Fanghao Hu is with the School of Electronic Information and Communications, Huazhong University of Science and Technology (HUST), Wuhan 430074, China (e-mail: \texttt{hfh@hust.edu.cn}).}
\thanks{Zhi Cai is with the School of Computer Science, Beijing University of Technology, Beijing 100124, China (e-mail: \texttt{caiz@bjut.edu.cn}).}
\thanks{Bang Wang is with the Hubei Key Laboratory of Smart Internet Technology, School of Electronic Information and Communications, Huazhong University of Science and Technology (HUST), Wuhan 430074, China (e-mail: \texttt{wangbang@hust.edu.cn}).}
\thanks{$^{*}$ denotes the corresponding author.}
\thanks{Manuscript received December 1, 2025.}}

\markboth{Journal of \LaTeX\ Class Files,~Vol.~14, No.~8, August~2021}%
{Shell \MakeLowercase{\textit{et al.}}: A Sample Article Using IEEEtran.cls for IEEE Journals}

\IEEEpubid{0000--0000/00\$00.00~\copyright~2021 IEEE}

\maketitle

\begin{abstract}
\textit{Community detection in multilayer networks} (CDMN) is to divide a set of entities with multiple relation types into a few disjoint subsets, which has many applications in the Web, transportation, and sociology systems. Recent neural network-based solutions to the CDMN task adopt a kind of \textit{representation fusion and global division} paradigm: Each node is first learned a kind of \textit{layer-wise representations} which are then fused for global community division. However, even with contrastive or attentive fusion mechanisms, the \textit{fused global representations} often lack the discriminative power to capture structural nuances unique to each layer.

\par
In this paper, we propose a novel paradigm for the CDMN task: \textbf{\underline{L}}ayered \textbf{\underline{D}}ivision and \textbf{\underline{G}}lobal \textbf{\underline{A}}llocation (\method{}). The core idea is to first perform layer-wise group division, based on which global community allocation is next performed. Concretely, \method{} employs a multi-head Transformer as the backbone representation encoder, where each head is for encoding node structural characteristics in each network layer. We integrate the Transformer with a community-latent encoder to capture community prototypes in each layer. A shared scorer performs layered division by generating layer-wise soft assignments, while global allocation assigns each node the community label with highest confidence across all layers to form the final consensus partition. We design a loss function that couples differentiable multilayer modularity with a cluster balance regularizer to train our model in an unsupervised manner. Extensive experiments on synthetic and real-world multilayer networks demonstrate that our LDGA outperforms the state-of-the-art competitors in terms of higher detected community modularities. Our code with parameter settings and datasets are available at \url{https://anonymous.4open.science/r/LDGA-552B/}.

\end{abstract}

\begin{IEEEkeywords}
Multi-layer Networks, Graph Learning, Community Detection, Transformer Model.
\end{IEEEkeywords}

\section{Introduction}
\label{sec:intro}
\par
Community detection, the task of identifying groups of nodes that are densely connected and share structural similarities in a network, has long been a central focus of web mining and network science\cite{fortunato2010community}. It supports many applications in diverse domains, like biology, economics, and sociology\cite{javed2018community}. Traditional methods, such as modularity maximization\cite{GM}, label propagation\cite{LabelPropagation}, or graph embedding\cite{grover2016node2vec}, have achieved lots of successes for community detection in \emph{single-layer networks}, where only one type of relation is modeled and all edges are hence of the same type.

\begin{figure}[t!]
    \centering
    \includegraphics[width=\linewidth]{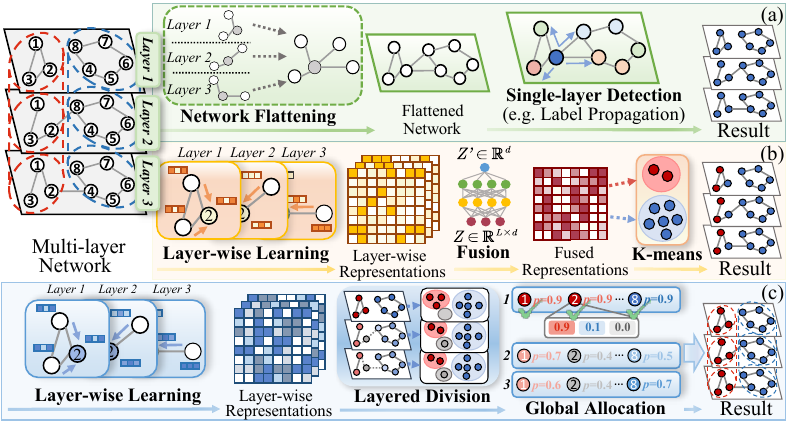}
    \vspace{-1.5em}
    \caption{Illustration of the two paradigms for community detection in multiplex networks. (a) and (b): representation fusion and global division (existing approaches); (c): layered division and global allocations (ours in this paper).}
    \label{fig:example}
    \vspace{-0.5em}
\end{figure}

\par
In reality, entities in Web systems and other domains often have multiple yet different types of relations. For example, users may be connected simultaneously through offline friendships, online hyperlinks, or financial transactions. Collapsing these heterogeneous interactions into a single layer discards crucial relational information and risks obscuring the true community structure. \emph{Multilayer networks} address this limitation by preserving all relational types, each by one network layer with edges representing one type of relation. Detecting communities in multilayer networks has numerous applications, such as uncovering the social groups of users on different online platforms~\cite{boccaletti2014structure}. 

\par
\IEEEpubidadjcol
Recently, many neural network-based solutions have been proposed for the community detection in multilayer networks~\cite{NFACC, MDeepWalk, MCLCD, NACC, GEAM, DMGI, HDMI}. However, they all follow a kind of \emph{representation fusion and global division} (\textbf{RFGD}) paradigm: neural encoders are designed to first learn layer-wise nodes' representations, that is, a representation is learned in each network layer for each node. Next, \emph{fusion mechanisms} are designed to fuse layer-wise representations into one global representation, that is, a global representation is fused for each node. Finally, clustering algorithms are designed to divide nodes into communities based on computing similarities between their \emph{global representations}.

\par
The RFGD paradigm has evolved significantly. Early strategies, as shown in Fig.~\ref{fig:example}(a), relied on network flattening , which collapses the multilayer structure into a single graph~\cite{berlingerio2013abacus}, so that standard detection algorithms (e.g., Label Propagation~\cite{LabelPropagation}, Louvain~\cite{Louvain} and Greedy Modularity~\cite{GM}) can be applied. In contrast, modern neural solutions are illustrated in \Cref{fig:example}(b). For example, MGCCN~\cite{MGCCN} evaluates the relevance coefficients of neighboring nodes for learning layer-wise representations; In order to fuse inter-layer semantic information, GEAM~\cite{GEAM} first applied a self-attention layer to learn layer-aware representations, then utilized an adaptive fusion mechanism to obtain fused representations. Then, the $K$-means algorithm is applied to cluster nodes into communities based on the fused representations. 

\begin{figure*}[t!]
    \centering
    \includegraphics[width=\textwidth]{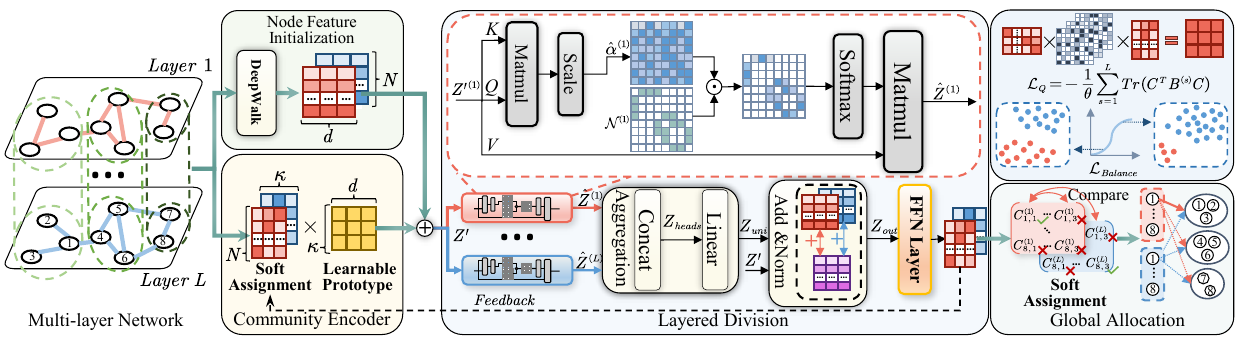}
    \vspace{-1.5em}
    \caption{The overview of \method{} framework (FFN denotes feed-forward networks). \method{}  first generates layer-wise soft assignments via Layered Division. In Global Allocation, it assigns each node its highest-confidence community across layers (e.g., node 1 to community 1). The model is trained end-to-end by maximizing a differentiable multilayer modularity objective.}
        \vspace{-0.5em}
    \label{fig:model}
\end{figure*}

\par
In this paper, we argue that such an RFGD paradigm might not be the best choice for the \textit{community detection in multilayer networks} (CDMN) task. 
It suffers from three key limitations. (1) \textbf{Layer fidelity.} Representation fusion may blur the distinctions of individual layers, either collapsing them into a single view or mixing them with concatenation/additive attention, which weakens the ability to attribute communities to specific relation types. (2) \textbf{Representation expressiveness.} Most recent approaches rely on the graph neural networks (e.g., GCN~\cite{kipf2016semi}/GAT~\cite{velickovic2017graph}) for per-layer encoding, but these models are prone to limited expressivity~\cite{xu2018powerful, egressy2024provably}, over-smoothing \cite{li2018deeper, song2023ordered}, and over-squashing~\cite{alon2020bottleneck}; as a result, they struggle to capture higher-order structural and long-range topological characteristics on each network layer (e.g., cycles \cite{chen2020can, chen2019equivalence}). More expressive, attention-based encoders (e.g., transformers \cite{maron2018invariant, kim2021transformers}) offer a stronger basis for layer-wise learning.  (3) \textbf{Objective mismatch.} Most methods optimize a kind of objective functions for reconstruction or contrastive surrogates, rather than the actual goal of community detection, leaving representation learning and community division decoupled. Together, these challenges call for a new paradigm that is more layer-faithful and representation-expressive, as well as a more task-aligned training objective for the CDMN task.

\par
To address these challenges, we introduce \textbf{\underline{L}}ayer \textbf{\underline{D}}ivision and \textbf{\underline{G}}lobal \textbf{\underline{A}}llocation framework (\textbf{\method{}}) for multilayer network community detection. {The core idea is
to first perform layer-wise group division, based on which global community allocation is next performed. Take the multilayer network in~\Cref{fig:example} as an example. For a pre-specified number of $\kappa=2$ communities, the layered division is to divide nodes into 2 layer-wise communities with probabilities. Node $v_{2}$ is assigned to the 1st-community in layer-1 with a probability of 0.9 (2nd-community in layer-1 with probability 0.1), 2nd-community in layer-2 with a probability of 0.4, and 2nd-community in layer-3 with a probability of 0.4. Then the global allocation will assign node $v_{2}$ to the global community-1, since it has the highest confidence across all layers.

\par
The \method{} employs a multi-head Transformer integrated with a layer-wise community-latent encoder as the backbone encoder. It first performs \textbf{layer-wise division}, where each network layer is encoded by a dedicated Transformer head with sparse dot-product attention restricted to observed edges. A \emph{community-latent encoder} (CLE) augments these per-layer representations with latent community prototypes, controlled by a stability-aware coefficient. An aggregation layer consists of concatenation and down-projection, which integrates the cross-layer signals to capture global structural contexts. A shared feed-forward scorer projects representations into layer-wise assignments. Then the \textbf{global allocation} mechanism compares the probabilities and selects labels with the highest confidence to obtain a single consensus partition.


Instead of optimizing a surrogate, \method{} is trained to directly maximize differentiable multilayer modularity, augmented with a balance regularizer that discourages degenerate solutions and behaves consistently across datasets. The model yields a consensus partition across layers by selecting labels with the highest confidence. This end‑to‑end objective couples attention, feature enrichment, and clustering, aligning learning signals with evaluation metrics.

Experimenting on synthetic mLFR benchmarks and ten real‑world multilayer networks spanning social, biological, and transportation domains, \method{} consistently improves community quality, raising NMI/ARI by 17.3\%/30.0\% on synthetic networks and multilayer modularity by 8.2\% on real-world networks, achieving the new state-of-the-art results. 



In summary, the contributions of this paper are threefold: 
\begin{itemize}[topsep=1pt,itemsep=1pt,parsep=0pt,leftmargin=10pt]
\item \textbf{Paradigm.} We propose a new \method{} paradigm, which divides nodes at each layer and allocates them to global communities. We implement the \method{} via a transformer with one head per layer as a backbone encoder integrated with a community encoder. 

\item \textbf{Objective.} We propose a differentiable multilayer modularity with a normalized balance regularizer that directly optimizes the community detection metric and produces a consensus partition.

\item \textbf{Evaluation.} We conduct extensive evaluations on both synthetic and real-world datasets, and results  demonstrate consistent gains over the state-of-the-art competitors.
\end{itemize}

\section{Related Works}
\label{sec:related_works}

This section briefly reviews the related works for community detection in multi-layer networks from two aspects: (1) the strategies employed for fusing information and detecting communities across different network layers, and (2) the learning paradigms used to derive community divisions. 

\textbf{\emph{Information Fusion and Community Detection in multi-layer Networks.}} A central challenge in multi-layer network analysis is integrating information across layers without losing critical structural details. Methodologies have evolved from direct structural aggregation to more nuanced, learning-based approaches. The most direct strategies involve "flattening" or aggregating the multi-layer graph into a single, weighted graph, upon which standard community detection algorithms can be applied~\cite{fortunato2010community}. For example, ABACUS~\cite{berlingerio2013abacus} sums the adjacency matrices of each layer; Cozzo et al. \cite{cozzo2016multilayer} proposed a method to construct a larger "supra-graph" that represents node-layer pairs. While computationally simple, these methods share a fundamental limitation: the premature compression of information via network flattening inevitably discards the unique topological details of each layer, potentially misleading the global community divisions.

\par
A more sophisticated RFGD paradigm has emerged for the CDMN task with the rise of \textit{graph neural networks} (GNNs)~\cite{kipf2016semi}. The general pipeline involves first learning layer-wise embeddings using methods like DeepWalk~\cite{perozzi2014deepwalk} and subsequently fusing them to produce a unified representation. In~\cite{GCFM}, a graph convolutional auto-encoder is proposed for learning a low-dimensional vector encoding for each node before clustering. More advanced models use attention mechanisms or network augmentation to learn a weighted combination of layer embeddings, as seen in models like GEAM\cite{GEAM}, NFACC\cite{NFACC}, and Kazim et al.~\cite{kazim2025multilayer}. However, even these sophisticated methods ultimately merge distinct layer representations into a single global representation for each node. This final fusion step can blur heterogeneous relational signals, losing the strict separation of layer-wise topologies.


\textbf{\emph{Learning Paradigms and Objective Functions for Graph Clustering.}} Beyond the fusion strategy, the learning objective function profoundly impacts the quality of the detected communities. The majority of recent deep learning methods follow a two-stage pipeline that separates representation learning from the final clustering step. In the first stage, a GNN encoder learns node embeddings in a self-supervised manner, often using contrastive learning frameworks like Deep Graph Infomax (DGI)~\cite{velivckovic2018deep}, DMGI~\cite{DMGI}, and NACC\cite{NACC}. In the second stage, a standard clustering algorithm like K-means is applied to cluster these embeddings. This decoupled pipeline, however, creates a significant objective mismatch. The GNN is trained to optimize a surrogate objective, such as a contrastive loss, which is not a direct measure of the clustering quality. An embedding optimized for general representation quality is not guaranteed to be optimal for partitioning the graph into dense communities.

\par
To resolve this mismatch, an end-to-end paradigm has emerged that directly optimizes a clustering-specific objective function. For single-layer networks, the recent work~\cite{tsitsulin2023graph} introduced a fully differentiable version of the classic modularity metric, allowing a GNN to be trained to directly maximize the clustering quality. Our proposed \method{} framework extends this powerful insight to multi-layer networks. By developing a differentiable multi-layer modularity loss, we align the entire learning process with the goal of community detection in multilayer networks, unifying feature learning and graph clustering into an end-to-end trainable system.

\section{Problem Definition}
\label{sec:problem_def}
\begin{definition}[Multi-layer Network]
    A multi-layer network is defined as a tuple $\mathcal{G}=\{\mathcal{V}, \mathcal{E}\}$, where $\mathcal{V}=\{v_1, v_2, \dots, v_N\}$ denotes the shared set of $N$ nodes across all layers. The edge set $\mathcal{E}=\{\mathcal{E}^{(1)}, \mathcal{E}^{(2)}, \dots, \mathcal{E}^{(L)}\}$ consists of $L$ independent edge sets, where $\mathcal{E}^{(l)}$ represents the intra-layer edges in the $l$-th layer. No inter-layer edges exist in this structure.
\end{definition}

\begin{definition}[Multi-layer Network Community Detection]
 In this work, we focus on disjoint community detection. Formally, disjoint multi-layer network community detection is defined as dividing the node set $\mathcal{V}$ into a partition $\mathcal{C} = \{C_1, C_2, \dots, C_{\kappa}\}$ satisfying two conditions:
\begin{enumerate}
    \item \textbf{Coverage:} $\bigcup_{k=1}^{\kappa} C_k = \mathcal{V}$;
    \item \textbf{Disjointness:} $C_i \cap C_j = \emptyset$ for all $i \neq j$.
\end{enumerate}
\end{definition}


\section{Methodology}
\label{sec:methods}
In this section, we present our \method{} framework, designed to jointly preserve layer fidelity, capture community-level semantics, and directly optimize for clustering quality. Our approach integrates three components: (1) a community-latent encoder (CLE) combined with a layer-aligned sparse transformer for layered division. (2) a global allocation mechanism that selects community labels with highest confidence to produce consensus partition, (3) an end-to-end clustering objective based on differentiable multilayer modularity.


As illustrated in \Cref{fig:model}, the general workflow is as follows.
First, per-layer node embeddings are generated from each graph layer to initialize representations. Next, these embeddings are enriched through the CLE and updated via layer-wise sparse attention, which propagates information only along observed edges within each layer.
The outputs are mapped into soft community assignments through a shared scorer that encourages a common semantic space across layers. Finally, the assignments are aggregated into a consensus partition, and the entire model is trained end-to-end by maximizing a differentiable multilayer modularity objective.


\subsection{Node Feature Initialization}

Our framework supports multilayer graphs with optional node attributes. If raw node features are provided, they can be directly incorporated as input to each layer. If no attributes are available, which is common in many benchmark datasets, we initialize features via an unsupervised embedding method such as DeepWalk~\cite{perozzi2014deepwalk}, applied independently to each layer. This produces per-layer embeddings \small{$\mZ^{(s)} = \left[\vz^{(s)}_1,\vz^{(s)}_2,\dots,\vz^{(s)}_n\right]^T\in \R^{N \times d}$} that capture local connectivity patterns while preserving layer identity. Stacking them yields a feature tensor $\mZ \in \R^{L \times N \times d}$.



\subsection{Community-Latent Encoder (CLE)}

Most existing deep learning methods focus on designing neural network architectures from the perspective of nodes and network layers~\cite{teng2024multi, GEAM, liu2024motif}, while overlooking community-level semantics. To enrich representations, we introduce a \textit{community-latent encoder} (CLE), which enriches node features with learnable prototypes that represent latent communities. For each layer $s$, CLE maintains a learnable prototype matrix $\mE^{(s)} \in \R^{\kappa\times d}$, where each row vector represents a learnable latent prototype for one community.

At iteration $t$, CLE augments the node features of layer $s$ using the soft assignments $\mC^{(s,t-1)} \in \R^{N\times \kappa}$ obtained in the previous iteration (detailed in \Cref{sec:ffn}). This design follows an expectation–maximization style refinement, where community memberships and prototypes are updated alternately. The enriched features at iteration $t$ are computed as
\vspace{-0.25em}
\begin{equation}
    \mZ'^{(s)}=\mZ^{(s)}+ \eta ^{(s)} \mC^{(s,t-1)} \mE^{(s)},
\end{equation}
where $\eta^{(s)}$ is a learnable fusion coefficient for each layer.  At the first iteration ($t=1$), we set $\mC^{(s,0)} = 0$ and initialize $\eta^{(s)} = 0$, ensuring that randomly initialized prototypes do not distort the base features $\mZ^{(s)}$. As training progresses, $\mC^{(s,t-1)}$ becomes more informative, $\eta^{(s)}$ increases adaptively, allowing CLE to gradually and stably infuse community-level semantics into the node features once both assignments and prototypes have become informative. 

\subsection{Layered Division}
Following the enrichment of node features by CLE, \method{} applies a \textit{layer-division mechanism} to preserve the unique topology of each layer. Each layer in a multilayer network encodes distinct structural information that should not be prematurely mixed. Therefore, to \textbf{preserve layer fidelity}, for a network with $L$ layers, we instantiate a transformer encoder with $L$ attention heads, where each head is exclusively dedicated to processing the information from one layer. This design ensures that information is propagated only within the observed topology of that layer, preventing cross-layer interference during representation learning.

\subsubsection{\textbf{Layer-aligned Sparse Transformer}}\label{sec:layer-aligned Transformer}
For a node $v_i$ and its neighbor \small{$v_j \!\in \!\mathcal{N}_i^{(s)}$} in layer $s$, the unnormalized attention score $e_{ij}^{(s)}$ is computed as:
\begin{equation}\scriptsize
    \vq_i^{(s)}=({\vz'}_i^{(s)})^T\mW_Q^{(s)},\ \vk_j^{(s)}=({\vz'}_j^{(s)})^T\mW_K^{(s)},\ e_{ij}^{(s)} = \frac{\vq_i^{(s)} (\vk_j^{(s)})^T}{\sqrt{d} }
\end{equation}
where ${\vz'}_i^{(s)}$ denotes the enriched feature of nodes $v_i$, $\mW_Q^{(s)}, \mW_K^{(s)}\in\R^{d \times d}$ are learnable projection matrices in head $s$. We then normalize these scores using a sparse softmax function defined over the neighborhood $\mathcal{N}_i^{(s)}$:
\begin{equation}
    \alpha^{(s)}_{ij} = \frac{\exp(e_{ij}^{(s)})}{\sum_{k \in \mathcal{N}_i^{(s)}} \exp(e_{ik}^{(s)})}
\end{equation}

This formulation ensures that attention is only computed between directly connected nodes, thereby grounding the model in the observed network structure. Importantly, it also yields computational efficiency: attention complexity becomes proportional to the number of observed edges $\sum_s |\mE^{(s)}|$ rather than the full $N^2$ possible pairs.


Through the $L$ attention head, the representation of node $v_i$ in layer $s$ is updated as $\hat{\vz}^{(s)}_i$ in \Cref{eq:updated}. Collecting all nodes yields the layer-wise output $\hat{\mZ}^{(s)}$, each capturing the topological structure of its corresponding layer. To integrate these signals, the layer-wise output will pass through an aggregation module consisting of a concatenation and a down-projection. As follows, concatenating across $L$ layers then produces the representations $\mZ_{\text{heads}} \in \R^{N \times (L\cdot d)}$, where $\mathbin\Vert$ denotes the concatenation operator:
\begin{equation}\scriptsize\label{eq:updated}
    \hat{\vz}^{(s)}_i = \underset{\forall j\in \mathcal{N}(i)}{\oplus}\left(\mAlpha^{(s)}_{ij}({\vz'}_j^{(s)})^T\mW_V^{(s)}\right), \ \mZ_{\text{heads}} = \underset{s \in [1, L]}{\mathbin\Vert}\left(\hat{\mZ}^{(s)}\right)
\end{equation}

The concatenated heads $\mZ_{\text{heads}}$ are projected through a learnable matrix $\mW_O\in \R^{(L\cdot d)\times d}$, yielding a unified representation $\mZ_{uni}$. Finally, this unified representation is element-wise added back to the original representation of each layer via a residual connection.
\begin{equation}
    \mZ_{\text{uni}} = \mZ_{\text{heads}}\mW_O, \quad\quad \mZ_{\text{out}}^{(s)} = \mZ_{\text{uni}}+\mZ'^{(s)}
\end{equation}

Here, $\mZ_{\text{uni}}$ aggregates global multi-layer information, while the residual term preserves the layer-wise features from CLE. This design ensures that each layer benefits from cross-layer context without losing its individual structural fidelity.


\subsubsection{\textbf{Feed-Forward Soft Assignment Scorer}}\label{sec:ffn}
After layer-specific attention, \method{} generates layered division probabilities by mapping each layer’s output into soft community assignments. For the $s$-th layer, a shared feed-forward scorer is applied:
\begin{equation}
\mC^{(s)}=\softmax\left(\mW_2\left(\sigma(\mW_{1}Z_{\text{out}}^{(s)}+b_1)\right)+b_2\right)
\end{equation}
where $\mW_1 \in \R^{d \times d_{\text{FFN}}}$, $\mW_2 \in \R^{d_{\text{FFN} \times \kappa}}$, $b_1 \in \R^{d_{\text{FFN}}}$, $b_2 \in \R^{\kappa}$, and $\sigma(\cdot)$ denotes a PReLU activation. This produces a soft assignment probability matrix $\mC^{(s)} \in \R^{N \times \kappa}$, where each row corresponds to a node’s distribution over $\kappa$ communities under layer $s$. Here, $\kappa$ is a hyperparameter representing the maximum number of communities pre-specified by the user. The same scorer is shared across all layers, ensuring that assignments from different layers are mapped into a common semantic space. Collectively, these form the layered division outputs $\left\{\mC^{(s)}\right\}_{s=1}^L$.




\subsection{Global Allocation}
Given the set of per-layer soft assignments $\left\{\mC^{(s)}\right\}^L_{s=1}$, the global allocation stage integrates the layer-wise assignments into a single, definitive community partition. Intuitively, some layers provide clearer signals than others for a node’s membership. To capture the most confident evidence, we aggregate across all layers and assign each node to the community with the highest probability:

\begin{equation}
    g_i=\underset{s\in[L],p\in[\kappa]}{\operatorname{argmax}}\left(\mC^{(s)}_{i,p}\right)
\end{equation}
where $C^{(s)}_{i,p}$ denotes the probability that node $v_i$ belongs to community $p$ in layer $s$, and $g_i \in \left\{1,\dots, \kappa\right\}$ is the resulting community label. This consensus step ensures that the final partition reflects the strongest structural evidence available, while preserving cross-layer consistency enforced by the shared scorer.

\subsection{Multi-layer Modularity Maximization Loss}
A central challenge in multilayer community detection is the design of an objective that aligns directly with the goal of community detection. Many contemporary methods rely on contrastive objectives or supervised proxies, where embeddings are first learned and then clustered by a separate algorithm (e.g., k-means, Louvain). This two-stage process is not end-to-end and does not directly optimize the community structure of interest.


Inspired by Tsitsulin et al.\cite{tsitsulin2023graph}, who introduced a differentiable relaxation of the Newman–Girvan modularity, we extend this idea to multilayer network. We propose a loss function, $\gL_\text{soft}$, which couples \emph{differentiable multilayer modularity} with a normalized \emph{cluster balance regularizer}:
\begin{equation}
\resizebox{0.9\columnwidth}{!}{$ 
\gL_{\text{soft}} = 
\underbrace{-\frac{1}{\theta}\sum_{s=1}^{L}\text{Tr}\!\big(\mC^\top \mB^{(s)} \mC\big)}_{\textstyle \gL_Q}
+
\underbrace{\alpha \cdot \frac{\kappa}{N^2 (\kappa-1)}\sum_{p=1}^{\kappa}\Big(\sum_{i=1}^N \mC_{i,p}-\tfrac{N}{\kappa}\Big)^2}_{\textstyle \gL_{\text{Balance}}}
$}
\end{equation}

\noindent
\textbf{\emph{Differentiable multilayer modularity ($\gL_Q$).}}
Here $\mC \in \mathbb{R}^{N \times \kappa}$ is the soft assignment matrix output by \method{}, and
$
\mB^{(s)} = \mA^{(s)} - \gamma_s \frac{d^{(s)} (d^{(s)})^\top}{2m_s}
$
is the modularity matrix of layer $s$ with degree vector $d^{(s)}$ and edge count $m_s$. Resolution parameter $\gamma_s$  controls the scale of communities in layer $s$. The normalization constant $\theta = \sum_{s=1}^L m_s$ represents the total number of edges. This term encourages partitions with dense intra-community edges and sparse inter-community edges consistently across layers. Since modularity is to be maximized, we subtract this term from the loss.


\noindent
\textbf{\emph{Cluster balance regularizer ($\gL_{\text{Balance}}$).}}
Optimizing modularity alone risks degenerate solutions where all nodes collapse into a single cluster. To avoid this, we introduce a balance penalty that discourages extreme imbalance in community sizes. The normalization constant $\tfrac{N^2(\kappa-1)}{\kappa}$ (derived below) scales this term into $[0,1]$, making the hyperparameter $\alpha$ more interpretable.

\subsubsection{Derivation of $\textstyle \gL_Q$}
The standard modularity $Q_{m}$ for a multilayer network is defined as:
\begin{align*}
Q_m &= \frac{1}{2\omega }\sum_{ijsr} \bigg[ \underbrace{(A_{ijs}-\gamma _s\frac{d_{is}d_{jr}}{2m_s})\delta (s,r)\delta (g_{is},g_{jr})}_{\text{Intra-layer structure}} \\
&\quad + \underbrace{\delta (i,j)\ell _{jsr}\delta (g_{is},g_{jr})}_{\text{Inter-layer coupling}} \bigg]
\end{align*}
where $A_{ijs}$ is the adjacency of node $i$ to $j$ in layer $s$ and $\delta(g_{is}, g_{jr})$ is 1 if nodes belong to the same community, and 0 otherwise. $\ell_{jsr}$ denotes the coupling parameter. $\omega$ is the total multilayer edge strength, defined as $\omega=(\sum_{ijs}A_{ijs}+\sum _{jsr}\ell _{jsr} /2)$. $m_s$ is the total intra-layer edge strength of the $s$th layer, defined as $m_s=(\sum_{ij}A_{ijs})/2$.

Since our goal is to find a single consensus partition for all layers (i.e., $g_{is} = g_i$ for all $s$), the inter-layer coupling term becomes redundant. Hence, the objective function simplifies to maximizing the sum of intra-layer modularities:
\begin{equation}
    Q_{simple} \propto \sum_{s=1}^{L} \sum_{i,j} \left( A_{ijs} - \gamma_s \frac{d_{is}d_{js}}{2m_s} \right) \delta(g_i, g_j).
\end{equation}

To enable end-to-end training, we relax the discrete Kronecker delta $\delta(g_i, g_j)$ into continuous soft assignments. Let $\mC \in \mathbb{R}^{N \times \kappa}$ be the soft assignment matrix. The probability that nodes $i$ and $j$ interact within the same community is approximated by the inner product of their assignment vectors: $\delta(g_i, g_j) \approx (\mC \mC^\top)_{ij}$.

Substituting this into the simplified modularity equation and rewriting in matrix trace form:
\begin{equation}
    \sum_{s=1}^{L} \sum_{i,j} \mB^{(s)}_{ij} (\mC \mC^\top)_{ij} = \sum_{s=1}^{L} \text{Tr}(\mC^\top \mB^{(s)} \mC).
\end{equation}
Incorporating the normalization constant $\theta$ yields $\gL_Q$.

\subsubsection{Derivation of the Normalization Constant}

Consider
\begin{equation}
S = \sum_{p=1}^{\kappa}\Big(\sum_{i=1}^{N}\mC_{i,p}-\tfrac{N}{\kappa}\Big)^2.
\end{equation}
The minimum of $S$ is 0 when all clusters are perfectly balanced. Since $S$ is a convex function (sum of squares) constrained by a fixed number $N$, its maximum is attained at the extreme boundaries of the solution space. Therefore, the maximum occurs when the distribution is most skewed (i.e., all nodes fall into one cluster). In that case:
\begin{equation}
S_{\max} = \Big(N-\tfrac{N}{\kappa}\Big)^2 + (\kappa-1)\Big(0-\tfrac{N}{\kappa}\Big)^2 
= \tfrac{N^2(\kappa-1)}{\kappa}.
\end{equation}
Thus the normalization constant is $\tfrac{N^2(\kappa-1)}{\kappa}$.

\subsubsection{Avoiding Trivial Solutions}

Our objective naturally rules out the trivial partition:
\begin{itemize}[topsep=1pt,itemsep=2pt,parsep=0pt,leftmargin=12pt]
\item \textbf{Modularity term:} For a single-cluster solution, $\gL_Q = 0$. Any non-trivial partition with positive modularity improves the loss.
\item \textbf{Balance term:} The trivial solution maximizes $\gL_{\text{Balance}}$. Any more balanced partition reduces the penalty.
\end{itemize}
Together, these effects guarantee that $\gL_{\text{soft}}$ prefers non-trivial partitions with positive modularity, driving the model toward meaningful community structures.

By unifying these components, our \method{} framework provides a principled solution to the key challenges of multilayer community detection. Specifically, it ensures that: (1) \textbf{layer fidelity is preserved} through the \method{} paradigm, which processes layers independently before deriving a global consensus, thus preventing the premature information fusion of prior methods; (2) \textbf{node representations are made more expressive and semantically rich}, as the layer-aligned transformer powerfully captures complex structural patterns while the CLE injects valuable community-level guidance; and (3) \textbf{the learning objective is directly aligned with the clustering goal}, as the entire model is trained end-to-end to maximize modularity, thereby resolving the critical objective mismatch inherent in the existing two-stage approaches.

\section{Experiments}
\label{sec:experiments}
\definecolor{first}{RGB}{0, 0, 243}
\definecolor{second}{RGB}{255, 192, 0}
\definecolor{third}{RGB}{91, 155, 213}
\newcommand{\res}[2]{#1 \textsubscript{$\pm$ #2}}
\newcommand{\fir}[1]{\textcolor{first}{\textbf{#1}}}

\par

\begin{table}[t!]
\centering
\small
\caption{Parameter settings of mLFR datasets}
\begin{tabular}{lll}
\toprule
Parameter & Description         & Value    \\
\midrule
d         & Average degree      & 16   \\
maxd      & Maximal degree      & 32   \\
$\kappa$          & Degree power-law    & 2    \\
$\psi$         & Community power-law & 1  \\
\bottomrule
\end{tabular}
\label{tab:mLFR}
\end{table}

\begin{table}[t!]
\centering
\small
\caption{Statistics of real-world datasets}
\begin{tabular}{l@{\hspace{0.4em}}c@{\hspace{0.4em}}c@{\hspace{0.4em}}c@{\hspace{0.4em}}c}
\toprule
Datasets   & Type           & \# Nodes & \# Edges & Layers \\
\midrule
AUCs       & Social         & 61    & 620   & 5      \\
CKM        & Social         & 246   & 1551  & 3      \\
Lazega     & Social         & 71    & 2223  & 3      \\
RM         & Social         & 94    & 1385  & 3      \\
Vickers    & Social         & 29    & 740   & 3      \\
C.Elegans  & Biology        & 279   & 5863  & 3      \\
Drosophila & Biology        & 8215  & 43366 & 7      \\
\small{SACCHPOMB}  & Biology        & 4092  & 63406 & 7      \\
Kapferer   & Finance        & 39    & 1018  & 4      \\
EUAir      & \small{Transportation} & 450   & 3588  & 37     \\
\bottomrule
\end{tabular}
\vspace{-1em}
\label{tab:stats}
\end{table}

\subsection{Experiment Setup}

\paragraph{Datasets.}

To evaluate the performance of our proposed method, we conduct experiments on both synthetic and real-world datasets. Synthetic datasets mainly refer to \dataset{mLFR}~\cite{mLFR}, which is a baseline model for generating multilayer networks. The mixing parameter $\mu$ in the mLFR datasets, indicating the probability of edges connecting to the nodes of the other communities, can be adjusted. A smaller $\mu$ indicates more obvious community structures, while a larger $\mu$ indicates more vague community structures. Table \ref{tab:mLFR} presents the detailed parameter settings of mLFR datasets in our experiment.

We also select ten extensive real-world multilayer networks datasets from domains covering society (\dataset{AUCs} \cite{AUCs}, \dataset{CKM} \cite{CKM}, \dataset{Lazega} \cite{lazega2001collegial, lazega2006}, \dataset{RM} \cite{RM}, \dataset{Vickers} \cite{vickers1981representing}), biology (\dataset{C.Elegans} \cite{CElegans}, \dataset{Drosophila} \cite{Drosophila, SACCH}, \dataset{SACCHPOMB} \cite{Drosophila, SACCH}), finance \dataset{Kapferer} \cite{kapferer}), transportation (\dataset{EUAir} \cite{EUair}). They are all collected manually and without ground-truth labels. Table \ref{tab:stats} presents their basic information.

\paragraph{Competitors.}
In order to verify the performance of our method, we select nine state-of-the-art community detection methods, which can be divided into the following categories. The first group is traditional community detection algorithms: GenLouvain(GL) \cite{GenLouvain}, M-Infomap(MI) \cite{MInfo}, MDLPA \cite{MDLPA}, MolTi \cite{MolTi}.
The second group is five multiplex embedding algorithms: HDMI \cite{HDMI}, DMGI \cite{DMGI}, GEAM \cite{GEAM}, NFACC \cite{NFACC}, NACC \cite{NACC}. 

\begin{table}[t!]
\centering
\scriptsize
\caption{Performance on mLFR synthetic datasets. We highlight the \fir{first} and \underline{second} best results.}
\begin{tabular}{l@{\hspace{0.6em}}l@{\hspace{0.6em}}l@{\hspace{0.6em}}l@{\hspace{0.6em}}l@{\hspace{0.6em}}l@{\hspace{0.6em}}l@{\hspace{0.6em}}l@{\hspace{0.6em}}l@{\hspace{0.6em}}l@{\hspace{0.6em}}l@{\hspace{0.6em}}l}
\toprule
\multicolumn{2}{c}{\rotatebox{55}{Metrics}} & \rotatebox{55}{GL} & \rotatebox{55}{MI}      & \rotatebox{55}{HDMI}    & \rotatebox{55}{DMGI}   & \rotatebox{55}{GEAM}            & \rotatebox{55}{MDLPA}  & \rotatebox{55}{MolTi}           & \rotatebox{55}{NFACC}         & \rotatebox{55}{NACC}    & \rotatebox{55}{\method{}}            \\ \midrule
\multirow{4}{*}{\rotatebox{90}{L=4, $\mu$=0.2}} & NMI     & 69.98  & 69.98  & 15.45  & 2.12 & \fir{100.0}      & 69.98 & \fir{100.0}      & \fir{100.0} & \fir{100.0}      & \fir{100.0}      \\
& ARI     & 50.53 & 50.53 & 6.70  & 1.82 & \fir{100.0}      & 50.53 & \fir{100.0}     & \fir{100.0}  & \fir{100.0}      & \fir{100.0}      \\
& Purity  & 73.63  & 73.63  & 66.05  & 66.05 & \fir{100.0}      & 73.63 & \fir{100.0}      & \fir{100.0}    & \fir{100.0}      & \fir{100.0}      \\
& Qm      & 30.91  & 30.91  & 11.63  & 8.23 & \fir{46.40} & 30.91 & \fir{46.40} & \fir{46.40} & \fir{46.40} & \fir{46.40} \\
\midrule
\multirow{4}{*}{\rotatebox{90}{L=4, $\mu$=0.3}} & NMI     & 0  & 2.26  & 10.72  & 11.76 & 32.63          & 0      & 87.90    & 91.83      & \underline{100.0}    & \fir{100.0} \\
& ARI     & 0 & -0.36 & 8.60  & 8.31 & 23.88          & 0      & 78.99          & 93.94 & \underline{97.95}    & \fir{100.0} \\
& Purtiy  & 63.61  & 63.61  & 63.61  & 63.61 & 63.61          & 63.61 & 90.80          & 97.19  & \underline{99.05}    & \fir{100.0} \\
& Qm      & 5.39  & 5.50  & 10.77  & 14.54 & 25.15          & 5.39 & \underline{35.66}    & 35.07   & \underline{35.66}    & \fir{35.95} \\ 
\midrule
\multirow{4}{*}{\rotatebox{90}{L=4, $\mu$=0.4}} & NMI     & 0  & 0  & 2.41  & 4.91 & 0.22          & 0      & \underline{55.18}    & 2.40  & 38.93          & \fir{88.85} \\
& ARI     & 0  & 0 & -1.02 & 4.85 & -0.16         & 0      & \underline{42.00}    & -2.40   & 31.30          & \fir{90.73} \\
& Purity  & 65.81  & 65.81  & 65.81  & 65.81 & 65.81          & 65.81 & \underline{77.18}    & 65.81    & 69.19          & \fir{95.84} \\
& Qm      & 5.40  & 5.40  & 8.29  & 10.18 & 21.76          & 5.40 & \underline{27.94}    & 21.99    & 25.44          & \fir{28.35} \\ 
\midrule
\multirow{4}{*}{\rotatebox{90}{L=8, $\mu$=0.2}} & NMI     & 0  & 78.82  & 12.89  & 8.27 & 67.21          & 73.24 & \fir{100.0}      & \fir{100.0}   & \fir{100.0}      & \fir{100.0}      \\
& ARI     & 0 & 84.97 & 9.58  & 7.72 & 62.98          & 56.88 & \fir{100.0}      & \fir{100.0}   & \fir{100.0}      & \fir{100.0}      \\
& Purity  & 66.56  & 93.63  & 66.56  & 66.56 & 82.86          & 77.61 & \fir{100.0}      & \fir{100.0}  & \fir{100.0}      & \fir{100.0}      \\
& Qm      & 5.60  & 42.01  & 10.32  & 11.24 & 37.47          & 33.31 & \fir{47.01} & \fir{47.01}    & \fir{47.01} & \fir{47.01} \\ 
\midrule
\multirow{4}{*}{\rotatebox{90}{L=8, $\mu$=0.3}} & NMI     & 0  & 10.83  & 11.52  & 11.68 & 22.64          & 0      & \underline{89.42}    & 51.58   & 77.28          & \fir{100.0}      \\ 
& ARI     & 0 & -0.30 & 5.96  & 12.27 & 7.97          & 0      & \underline{83.59}    & 38.18   & 75.28          & \fir{100.0}      \\ 
& Purity  & 65.92  & 65.92  & 65.92  & 65.92 & 65.92          & 65.92 & \underline{92.99}    & 72.17    & 88.77          & \fir{100.0}      \\ 
& Qm      & 5.56  & 6.11  & 11.87  & 12.92 & 24.28          & 5.56 & \underline{36.68}    & 30.90   & 34.04          & \fir{37.59} \\ 
\midrule
\multirow{4}{*}{\rotatebox{90}{L=8, $\mu$=0.4}} & NMI     & 0  & 10.94  & 0.73  & 11.42 & 0.81          & 0      & \underline{40.87}    & 1.78 & 28.01          & \fir{52.18} \\ 
& ARI     & 0 & 0.84 & -0.54 & 8.08 & 0.51          & 0      & \underline{26.50}    & 1.35 & 18.68          & \fir{37.74} \\ 
& Purity  & 65.76  & 65.76  & 65.76  & 65.76 & 65.76          & 65.76 & \underline{71.70}    & 65.76    & 65.76          & \fir{71.88} \\ 
& Qm      & 5.26  & 7.01  & 17.72  & 10.55 & 24.38          & 5.26 & 26.28         & 24.06   & \underline{26.37}    & \fir{28.75} \\ 
\bottomrule
\end{tabular}
\vspace{-1em}
\label{tab:mLFR_result}
\end{table}

\begin{figure}[t!]
    \centering
    \includegraphics[width=\linewidth]{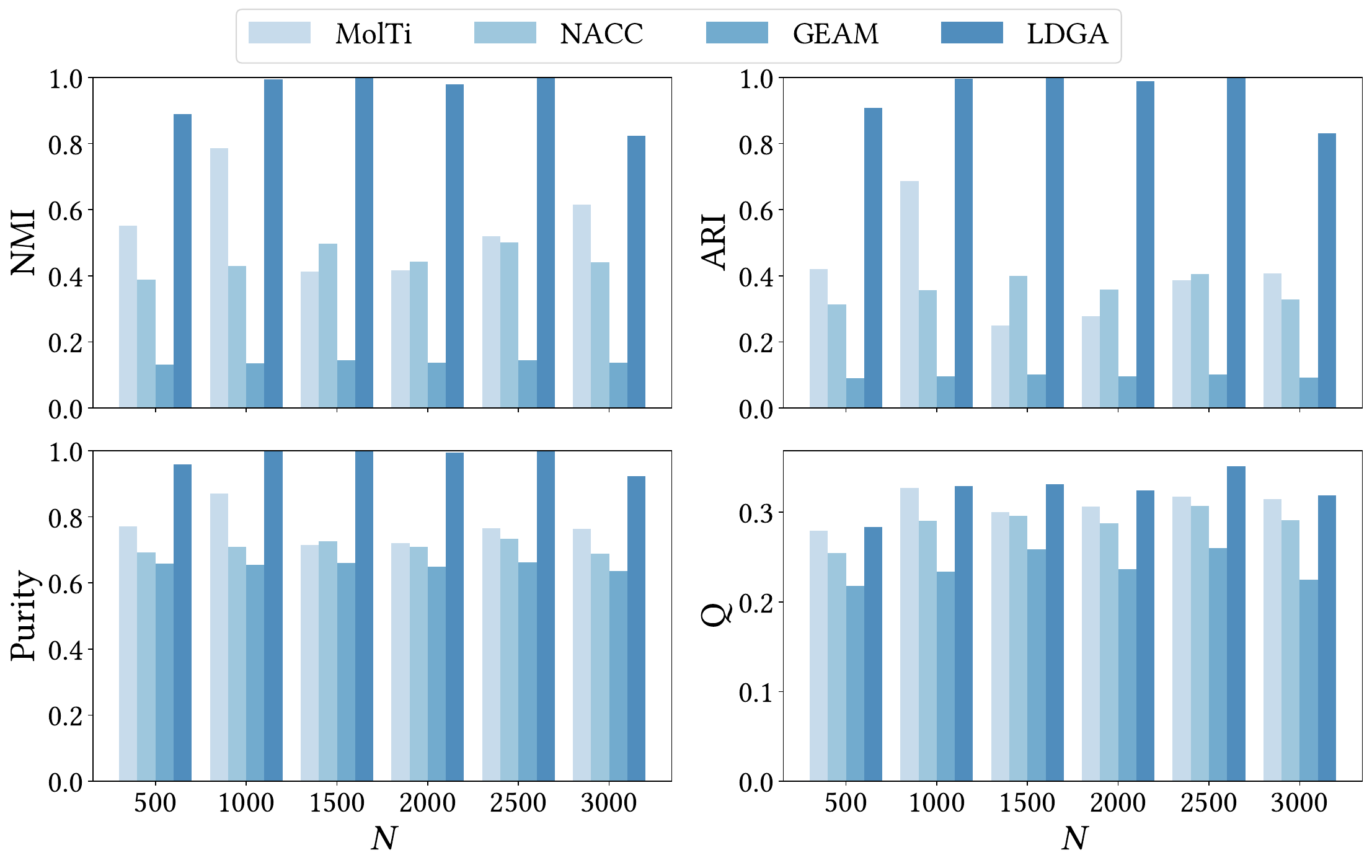}
    \vspace{-1.5em}
    \caption{Experiments with different $N$ on mLFR datasets, $N$ denotes the number of nodes in the network.}
    \vspace{-1.5em}
    \label{fig:mLFR_diffN}
\end{figure}

\paragraph{Evaluation.}
We use common evaluation metrics of community detection to evaluate the performance of all the methods, including NMI (Normalized Mutual Information)~\cite{NMI}, ARI (Adjusted Rand Index)~\cite{ARI}, Purity~\cite{Purity}, and multilayer modularity~\cite{mucha2010community}. The maximum value of multilayer modularity is 1, and a higher modularity means a clearer community structure. NMI, ARI, and Purity are only applied to synthetic datasets since they need ground-truth labels, while the modularity of detected communities is applied to both synthetic and real-world datasets. The resolution parameter and coupling parameter of modularity are both set to 1. For all experiments, we run 10 trials with different random seeds, and report the best results in Tables \ref{tab:mLFR_result}, \ref{tab:realworld}, and Fig. \ref{fig:mLFR_diffN}.

\paragraph{Parameters}
For the initial node feature, we generate a 512-dimensional vector for each network layer. For the DeepWalk, we set the window size as 10, the walks per node as 20 and the walk length as 80. The dimensions of the hidden layers in Transformer and FFN scorer are set to 512 and 1024 separately. We adopt the AdamW optimizer with the initial learning rate 0.0002.

For GL and MI, we use consensus voting to unify the community label. The resolution of GL is set to 1. For the rest of the methods, we follow the parameter settings recommended in their original papers or use the default values provided in their publicly available implementations to ensure a fair comparison. 

\begin{table*}[t!]
\centering
\caption{Modularity on real-world datasets. We highlight the \fir{first} and \underline{second} best results.}
\begin{tabular}{lllllllllll}
\toprule
\textbf{Network} & \textbf{GL} & \textbf{IM} & \textbf{HDMI} & \textbf{GEAM} & \textbf{DMGI}   & \textbf{MDLPA}    & \textbf{MolTi}   & \textbf{NFACC}   & \textbf{NACC} & \textbf{\method{}}      \\ \midrule
AUCs                                   & 43.80(7)        & 42.25(11)   & 39.54(11)    & 26.29(10)    & 23.33(3)       & 45.36(7)         & \underline{48.34(5)}  & 47.68(6) & 48.28(6)     & \fir{48.60(5)} \\
CKM                                    & 61.67(10)        & 56.55(29)   & 27.17(6)     & 53.66(4)     & 31.14(3)       & 47.81(46)        & \underline{61.71(5)}  & 53.23(6)    & 54.31(6)     & \fir{61.98(6)} \\
Lazega                                 & 6.96(1)        & 6.96(1)   & 17.65(12)    & 13.31(8)     & 08.29(3)       & 6.96(1)         & \underline{30.14(3)}  & 29.44(3) & 26.12(5)     & \fir{32.91(5)} \\
RM                     & 39.45(9)        & 32.55(15)   & 30.26(6)     & 25.38(3)     & 25.75(5)       & 43.22(3)         & \underline{43.57(5)} & 38.58(2)  & 30.32(5)     & \fir{45.15(5)} \\
Vickers                                & 9.89(2)        & 9.89(2)   & 27.18(3)     & 18.45(4)     & \underline{29.05(3)} & 22.24(2)         & 26.63(3)        & 28.72(3)    & 24.26(5)     & \fir{29.95(5)}   \\
C.Elegans                              & 25.49(8)        & 33.41(32)   & 17.97(11)    & 23.82(9)     & 15.53(4)       & 10.93(2)         & \underline{41.81(7)}  & 41.69(3) & 41.42(5)     & \fir{42.94(7)} \\
Drosophila                             & 30.22(139)       & \underline{33.27(845)}  & 15.44(3)     & 24.79(17)     & 18.35(5)       & 31.25(170) & Overflow.        & Overflow. & Overflow.     & \fir{40.13(5)} \\
SACCH.                             & 19.90(62)       & 18.75(627)  & 11.16(12)    & 19.05(9)     & 12.75(3)       & 14.38(74)        & \underline{24.68(30)} & 6.48(19) & 7.04(14)   & \fir{28.76(8)} \\
Kapferer                               & 13.58(6)        & 20.81(12)   & 17.37(8)     & 16.83(9)     & 20.81(3)       & 11.26(1)         & \underline{28.97(4)}  & 24.82(6) & 26.19(6)     & \fir{29.73(3)} \\
EUAir                                & 19.12(50)        & 17.94(175)   & 23.03(3)     & 24.15(3)     & 20.20(3)       & 18.67(19)        & \underline{24.71(12)} & 18.23(2)    & 15.28(10)    & \fir{30.37(6)} \\
\bottomrule
\end{tabular}
\label{tab:realworld}
\end{table*}

\begin{table*}[t!]
\centering
\caption{Results of ablation study. We highlight the \fir{first} and \underline{second} best results.}
\begin{tabular}{lccccccccccccc}
\toprule
Method     & \multicolumn{1}{l}{AUCs} & \multicolumn{1}{l}{C.Ele.} & \multicolumn{1}{l}{CKM} & \multicolumn{1}{l}{Dro.} & \multicolumn{1}{l}{EUAir} & \multicolumn{1}{l}{Kap.} & \multicolumn{1}{l}{Laz.} & \multicolumn{1}{l}{RM} & \multicolumn{1}{l}{SAC.} & \multicolumn{1}{l}{Vic.} & \multicolumn{1}{l}{SYN4-0.2} & \multicolumn{1}{l}{SYN4-0.3} & \multicolumn{1}{l}{SYN4-0.4} \\ \midrule
w/o Res.   & \underline{48.33}                   & \underline{42.78}                        & 59.27                  & 32.33                         & \underline{29.91}                      & \fir{29.73}                       & \fir{32.91}                     & 42.53                 & \underline{28.48}                        & \underline{29.14}                      & \fir{46.40}                         & \underline{35.87}                         & 23.27                         \\
w/o CLE    & \underline{48.33}                   & 42.61                        & \underline{61.66}                  & \underline{36.10}                         & 29.66                      & 27.95                       & \fir{32.91}                     & \underline{44.84}                 & 28.36                        & 29.07                      & \fir{46.40}                         & \underline{35.87}                         & \underline{24.04}                         \\
w/o Trans. & 41.03                   & 34.52                        & 56.75                  & 30.39                         & 24.60                      & 25.69                       & 26.98                     & 42.26                 & 17.63                        & 28.66                      & 46.10                         & 31.77                         & 20.06                         \\ \midrule
\method{}    & \fir{48.60}                   & \fir{42.94}                        & \fir{61.98}                  & \fir{40.13}                         & \fir{30.37}                      & \fir{29.73}                       & \fir{32.91}                     & \fir{45.15}                 & \fir{28.76}                        & \fir{29.95}                      & \fir{46.40}                         & \fir{35.95}                         & \fir{28.35}                    \\ \bottomrule       
\end{tabular}
\label{tab:ablation}
\end{table*}

\subsection{Experimental Results}


\par
\emph{\textbf{Q1: How effectively does \method{} identify controllable community structures in synthetic networks?}}
On \dataset{mLFR} datasets (\Cref{tab:mLFR_result}), \method{} consistently and significantly outperforms all baseline across different numbers of layers($L=4,8$) and mixing parameters ($\mu=0.2,0.3,0.4$). The gains are especially pronounced in challenging settings with high inter-community mixing. For instance, at $L=4$ and $\mu=0.4$, \method{} achieves an NMI of 88.85\%, whereas the next-best performing method (MolTi), reaches only 55.18\%. This shows that \method{} maintains strong discriminative power even when community boundaries are blurred. In contrast, classical multi-layer heuristics (e.g., GL) fail to capture meaningful communities, and deep learning baselines such as DMGI and GEAM lag due to their reliance on surrogate objectives and two-stage clustering pipelines. 


\par
\Cref{fig:mLFR_diffN} presents the scalability of the top-performing methods as the number of nodes ($N$) increases from 500 to 3000. \method{} not only achieves the best scores overall but also maintains near-perfect NMI, ARI, and Purity as the graph grows, demonstrating superior stability compared to MolTi and NACC, whose performance either fluctuates or declines. This scalability stems from \method{}’s sparse attention, whose complexity scales with the number of edges rather than quadratically with nodes, making it well-suited for large-scale multilayer networks.



\par
\emph{\textbf{Q2: How well does \method{} generalize across diverse, real-world application domains?}}
\method{} achieves the highest multilayer modularity on all ten real-world datasets  (\Cref{tab:realworld}), demonstrating strong generalizability across domains, including social (CKM, Lazega), biological (C. Elegans, SACCHPOMB), and transportation (EUAir). For instance, on EUAir, \method{} reaches a modularity of 30.37\%, compared to 24.71\% from the strongest baseline (MolTi). On the large-scale Drosophila network, where several deep learning methods failed due to memory constraints, \method{} not only remained scalable but also achieved the top score of 40.13\%. These results confirm that \method{} consistently discovers higher-quality communities and that its end-to-end modularity-driven optimization is effective for uncovering potential community structure in complex, heterogeneous systems.


\subsection{Ablation Study}
\textit{\textbf{Q3: What are the contributions of \method{}'s core components to its overall performance?}}
The ablation study reported in \Cref{tab:ablation} shows that all core components (Transformer, CLE, and residual connections) are integral and work synergistically to achieve \method{}’s success. We systematically removed each component and evaluated the resulting modularity on both synthetic and real-world datasets (e.g., SYN4-0.2 denotes an mLFR network with 500 nodes, 4 layers, and mixing parameter $\mu=0.2$. (1) \textbf{Removing Transformer module ("w/o Trans.") resulted in the most significant performance drop.} For instance, on CKM, the modularity decreased from 61.98\% to 56.75\%, and on SYN4-0.4, the modularity plummeted from 35.95\% to 31.77\%. This highlights the Transformer's crucial role in effectively integrating multilayer information and learning robust cross-layer node representations. 
(2) \textbf{The absence of the CLE ("w/o CLE") also led to a noticeable performance decline.} On Kapferer, the modularity dropped from 29.73\% to 27.95\%, and on SYN4-0.4, modularity decreased to 24.04\%. This suggests that injecting learnable community prototypes enhances the model's ability to discriminate between communities, providing valuable semantic guidance.
(3) \textbf{The removal of the residual connection caused a slight but consistent decrease in modularity.} For example, on CKM, the modularity reduced to 59.27\%, and on Drosophila, it became 32.33\%. While less dramatic, this indicates that direct preservation of initial layer-wise features, alongside globally aggregated information, is beneficial for maintaining the quality and stability of learned embeddings.

In summary, these results highlight that each module contributes uniquely, and their combination is essential for \method{}’s state-of-the-art performance in multilayer network community detection.

\begin{figure*}[!t]
    \centering
    \includegraphics[width=\textwidth, trim=0cm 0cm 0cm 0cm, clip]{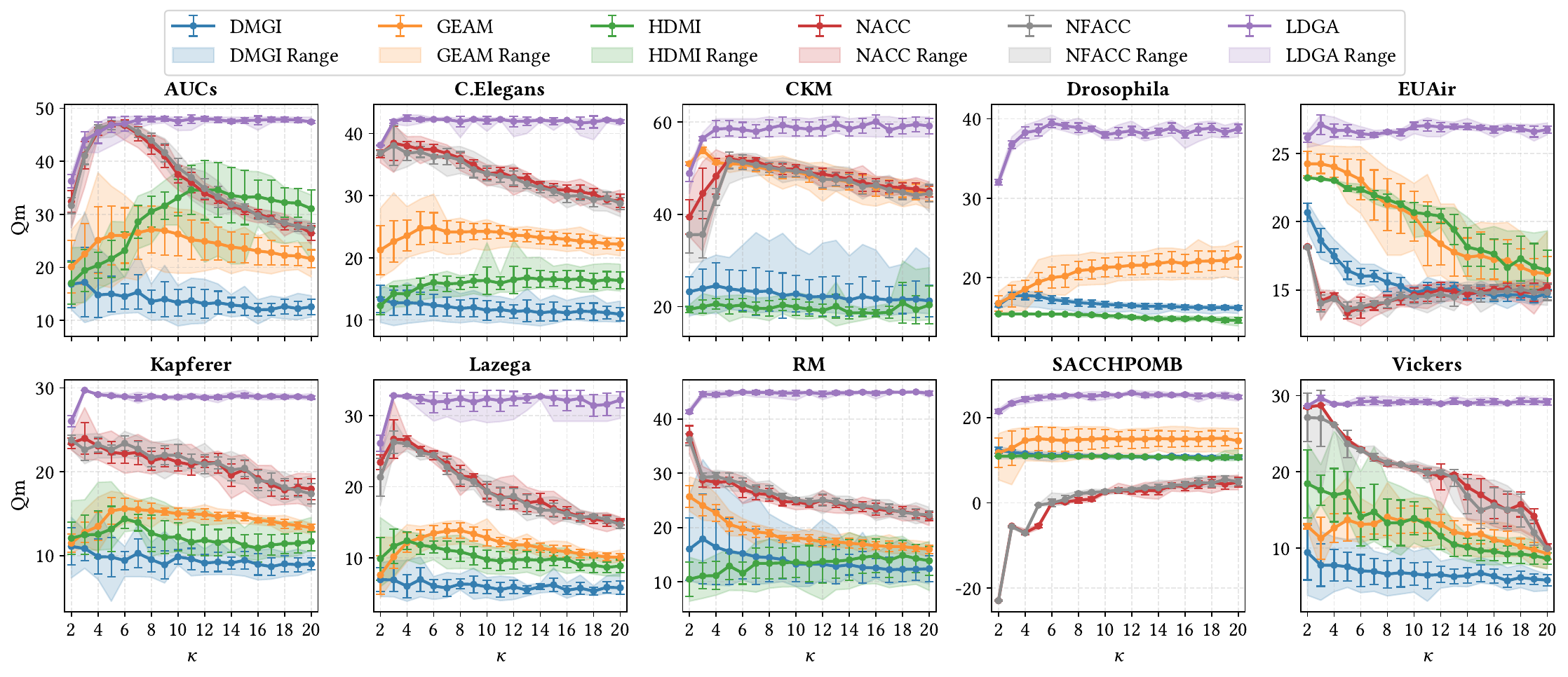}
    \caption{Modularity with different pre-specified maximum numbers of communities $\kappa$ on ten real-world datasets.}
    \label{fig:diffK}
\end{figure*}

\subsection{Hyperparameter Sensitivity Analysis}
\emph{\textbf{Q4: How sensitive is \method{} to $\kappa$, the pre-specified maximum number of communities, a common challenge in real-world applications where ground truth is unknown?}}
\method{} exhibits strong robustness to the hyperparameter $\kappa$, a hyperparameter represents the maximum number of communities. In practice, the true number of communities is rarely known, making robustness to hyperparameter $\kappa$ critical. \Cref{fig:diffK} plots the modularity achieved by \method{} and four leading deep-learning based baselines (DMGI, GEAM, HDMI, NACC) as $\kappa$ varies from 2 to 20. For each value of $\kappa$, results are averaged over 10 runs with different random seeds, with shaded bands indicating variance.

\par
Across most datasets (e.g., Kapferer, Lazega, RM), \method{} maintains consistently high modularity regardless of $\kappa$. In contrast, baselines are highly sensitive: their scores peak sharply at particular values of $\kappa$ and degrade rapidly elsewhere. This stability highlights a key practical advantage of \method{}—its partitions are guided by the intrinsic multilayer structure rather than overfitting to the pre-specified $\kappa$, making it reliable even when the number of communities is misspecified.

\begin{figure}[t!]
    \centering
    \includegraphics[width=\linewidth]{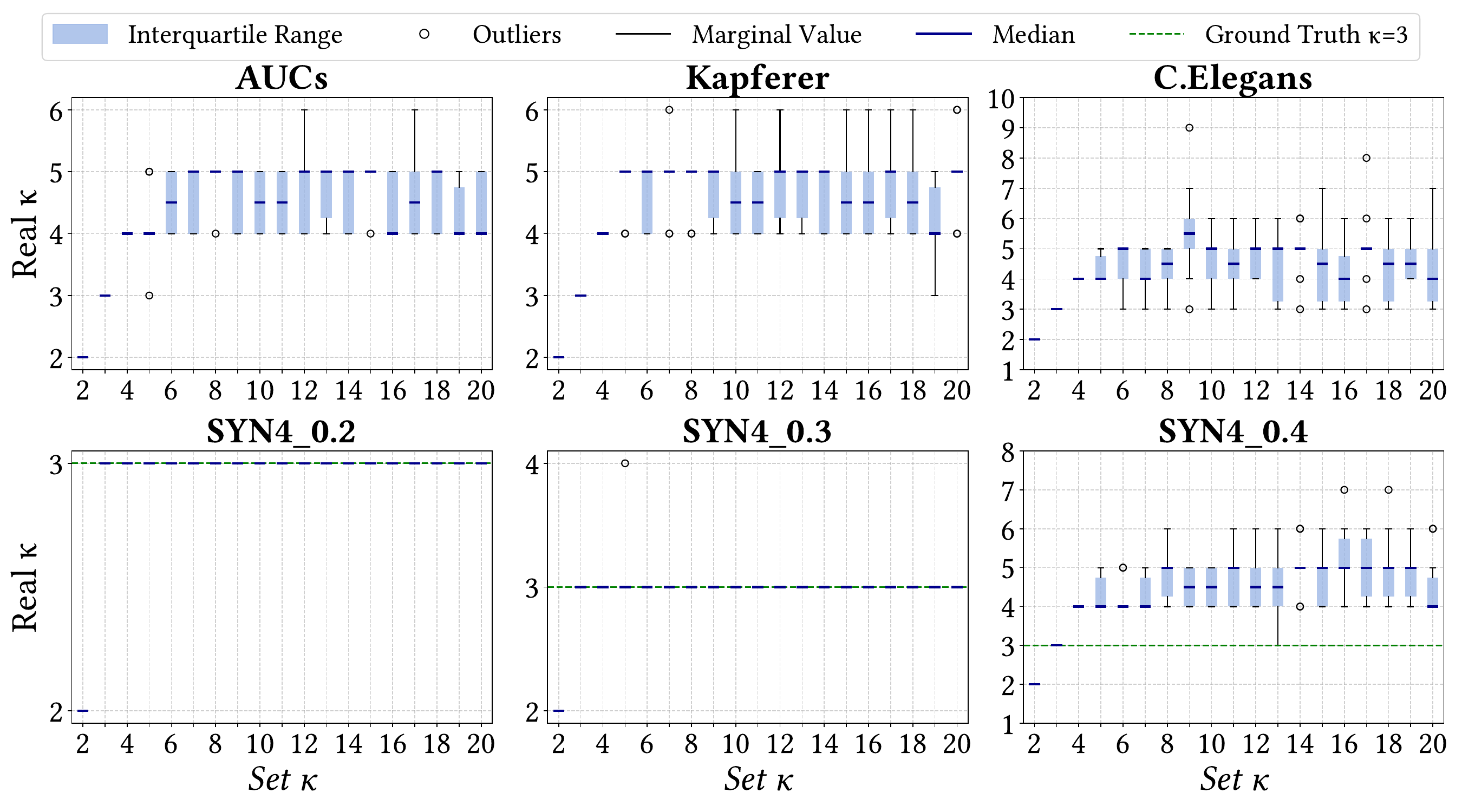}
    \vspace{-2em}
    \caption{Real community amount on different datasets}
    \vspace{-1em}
    \label{fig:setKandrealK}
\end{figure}

\emph{\textbf{Q5: What mechanism underlies \method{}'s robustness, and can it automatically determine an appropriate number of communities?}}
\method{}’s robustness arises from its ability to discover the most natural partition of a network. The number of non-empty communities it produces quickly converges to a stable, narrow range that is independent of \mbox{the pre-specified $\kappa$}.

To investigate this mechanism, we analyzed the relationship between the pre-specified maximum number of communities (“Set $\kappa$”) and the actual number of non-empty communities produced by the model (“Real $\kappa$”). Figure~\ref{fig:setKandrealK} illustrates this relationship on both real-world and synthetic datasets. The results of NACC and NFACC in Drosophila dataset are missing because of memory overflow.

On real-world networks (top row), “Real $\kappa$” does not grow unbounded as “Set $\kappa$” increases. Instead, it stabilizes within a narrow range. For example, on AUCs, the detected communities consistently converge between 5 and 7, regardless of the chosen “Set $\kappa$.” This indicates that \method{} captures the inherent community structure rather than forcing arbitrary partitions.

The effect is clearer on mLFR datasets (bottom row), where the ground-truth community number is $\kappa=3$ (green dashed line). Even when “Set $\kappa$” increases well beyond this value, “Real $\kappa$” remains tightly centered around 3. This provides strong evidence that the modularity-driven objective effectively prunes redundant or empty clusters, guiding the model toward the most natural partition.

In summary, \method{} not only maintains stable performance across different $\kappa$ values but also demonstrates an inherent ability to infer an appropriate number of communities—reliably uncovering the latent structure of multilayer networks.
\begin{table*}[th]
\centering
\caption{Running time(s) on real-world datasets in 100 epochs. We highlight the \fir{first} and \underline{second} maximum value.}
\begin{tabular}{c|l|llllllllll}
\toprule
\multicolumn{1}{l}{}                     & \multicolumn{1}{l}{Method} & AUCs            & C.Ele.          & CKM             & Dro.             & EUAir           & Kap.            & Laz.            & SAC.             & RM              & Vic.            \\ \midrule
\multirow{6}{*}{AVG Time(in 100 epochs)} & DMGI                       & \underline{0.0976}    & \underline{0.0994}    & \underline{0.0933}    & 0.1231           & \underline{0.9178}    & \fir{0.1456} & \fir{0.1343} & 0.1308           & \underline{0.0468}    & \fir{0.1106} \\
                                         & HDMI                       & 0.0341          & 0.0222          & 0.0212          & 0.1505           & 0.2115          & 0.0254          & 0.0213          & 0.0901           & 0.0201          & 0.0229          \\
                                         & NFACC  & 0.0118          & 0.0101          & 0.0324          & Overflow.        & 0.0710          & 0.0079          & 0.0071          & 1.8646           & 0.0068          & 0.0067          \\
                                         & NACC                       & 0.0168          & 0.0071          & 0.0069          & Overflow.        & 0.0388          & 0.0081          & 0.0058          & \underline{2.6970}     & 0.0053          & 0.0057          \\
                                         & GEAM                       & \fir{0.2160} & \fir{0.5085} & \fir{0.4504} & \fir{71.5126} & \fir{6.5550} & \underline{0.1186}    & \underline{0.1232}    & \fir{18.4292} & \fir{0.1478} & \underline{0.0886}    \\
                                         & Ours                       & 0.0264          & 0.0162          & 0.0166          & \underline{0.3139}     & 0.2464          & 0.0195          & 0.0153          & 0.1413           & 0.0135          & 0.0171          \\ \midrule
\multirow{6}{*}{Max Time(in 100 epochs)} & DMGI                       & \underline{1.0504}    & \underline{0.2071}    & \underline{0.1877}    & 0.3574           & \underline{3.1314}    & \underline{0.1934}    & \underline{0.1662}    & 0.4942           & \underline{0.0747}    & \underline{0.1679}    \\
                                         & HDMI                       & 1.3397          & 0.0442          & 0.0339          & \underline{0.4568}     & 0.3320          & 0.0379          & 0.0355          & 0.3212           & 0.0303          & 0.0317          \\
                                         & NFACC  & 0.1571          & 0.0217          & 0.0598          & Overflow.        & 0.1967          & 0.0197          & 0.0124          & 3.6519           & 0.0115          & 0.0201          \\
                                         & NACC                       & 0.8198          & 0.0258          & 0.0174          & Overflow.        & 0.1321          & 0.0185          & 0.0133          & \underline{4.8516}     & 0.0075          & 0.0190          \\
                                         & GEAM                       & \fir{2.7211} & \fir{0.8479} & \fir{0.7524} & \fir{94.4320} & \fir{6.7426} & \fir{0.3025} & \fir{0.3098} & \fir{20.3191} & \fir{0.3258} & \fir{0.2758} \\
                                         & Ours                       & 0.4363          & 0.0348          & 0.0360          & 0.4369           & 0.3354          & 0.0457          & 0.0315          & 0.1658           & 0.0277          & 0.0508          \\ \midrule
\multirow{6}{*}{Min Time(in 100 epochs)}      & DMGI                       & \underline{0.0607}    & \underline{0.0478}    & \underline{0.0462}    & 0.1028           & \underline{0.4777}    & \underline{0.0576}    & \fir{0.1086} & 0.1137           & \underline{0.0415}    & \underline{0.0479}    \\
                                         & HDMI                       & 0.0267          & 0.0183          & 0.0183          & 0.1457           & 0.1919          & 0.0221          & 0.0178          & 0.0859           & 0.0177          & 0.0182          \\
                                         & NFACC  & 0.0077          & 0.0059          & 0.0122          & Overflow.        & 0.0477          & 0.0066          & 0.0051          & 0.0128           & 0.0040          & 0.0040          \\
                                         & NACC                       & 0.0055          & 0.0045          & 0.0044          & Overflow.        & 0.0309          & 0.0051          & 0.0040          & \underline{1.5038}     & 0.0041          & 0.0040          \\
                                         & GEAM                       & \fir{0.1412} & \fir{0.4026} & \fir{0.3861} & \fir{51.6594} & \fir{6.3588} & \fir{0.0934} & \underline{0.0991}    & \fir{18.0090} & \fir{0.1163} & \fir{0.0567} \\
                                         & Ours                       & 0.0182          & 0.0128          & 0.0120          & \underline{0.2794}     & 0.2102          & 0.0152          & 0.0116          & 0.1270           & 0.0113          & 0.0116          \\ \midrule
\multirow{6}{*}{Std(in 100 epochs)}      & DMGI                       & \underline{0.1770}    & \underline{0.0420}    & \underline{0.0413}    & \underline{0.0176}     & \fir{0.3911} & \underline{0.0454}    & \underline{0.0105}    & 0.0530           & \underline{0.0055}    & \underline{0.0396}    \\
                                         & HDMI                       & 0.0435          & 0.0031          & 0.0021          & 0.0164           & 0.0189          & 0.0022          & 0.0026          & 0.0150           & 0.0019          & 0.0027          \\
                                         & NFACC  & 0.0155          & 0.0047          & 0.0132          & Overflow.        & 0.0330          & 0.0018          & 0.0016          & \fir{0.8183}  & 0.0014          & 0.0021          \\
                                         & NACC                       & 0.0852          & 0.0028          & 0.0019          & Overflow.        & 0.0133          & 0.0024          & 0.0018          & \underline{0.7370}     & 0.0007          & 0.0020          \\
                                         & GEAM                       & \fir{0.2590} & \fir{0.0986} & \fir{0.0805} & \fir{14.3347} & \underline{0.1133}    & \fir{0.0584} & \fir{0.5708} & 0.4689           & \fir{0.0565} & \fir{0.0593} \\
                                         & Ours                       & 0.0412          & 0.0030          & 0.0038          & 0.0151           & 0.0231          & 0.0039          & 0.0032          & 0.0070           & 0.0025          & 0.0078         \\ \bottomrule
\end{tabular}
\label{Tab:Time}
\end{table*}

\subsection{Time Analysis}
\emph{\textbf{Q6: How is the computational efficiency of \method{}?}} We conducted a time-latency analysis, comparing \method{}'s performance against leading deep learning-based baselines. The experiments were performed on ten real-world datasets, measuring the wall-clock time required to complete 100 training epochs. The results(Table \ref{Tab:Time}), demonstrate that \method{} achieves a balance between high performance and computational efficiency.

Our model consistently operates on par with, or faster than, most baseline methods while maintaining high community detection quality. For instance, on the CKM dataset, \method{}'s average runtime is 0.0166 seconds per epoch, significantly faster than competitors like GEAM (0.4504s), DMGI (0.0933s), and HDMI (0.0212s). On the largest dataset Drosophila, \method{}'s running time (0.3139s) remains the same order of magnitude as the fastest method (0.1231s).

The computational advantage of \method{} is primarily attributed to its layer-aligned sparse transformer architecture. As discussed in Section \ref{sec:layer-aligned Transformer}, the sparse attention mechanism restricts computations to observed edges only. This design choice avoids the quadratic complexity ($O(N^2)$) associated with standard transformers, making the model's complexity proportional to the number of edges ($O\sum_s|E^{(s)}|$). This ensures that \method{} scales effectively with network size and sparsity, positioning it as a practical and scalable solution for community detection in large, real-world multilayer systems.

\section{Conclusion}
\label{sec:Conclusion}
\par
In this paper, we have proposed \method{}, a novel paradigm for multilayer community detection. The \method{} first performs layered division to generate layer-wise soft assignments, then globally allocates each node to a community with the highest confidence across all layers to form a consensus partition. To achieve better feature expressiveness, we have designed a community-latent encoder into a layer-aligned transformer. A differentiable multilayer modularity loss is designed to align the training process with the community detection task. Experiments on synthetic and real-world networks show consistent improvements in our \method{} over competitors. This work highlights the promise of end-to-end, modularity-driven approaches for scalable community detection in Web-scale multilayer systems.

\par
Our framework is still limited in the attention mechanism restricted to observed edges. Future work shall focus on designing a more expressive attention mechanism to capture long-range structural patterns.

%

\bibliographystyle{IEEEtran}
\bibliography{IEEEabrv}


 




\newpage
\appendix[Additional Experiment Details]
\label{sec:appendix}
\definecolor{first}{RGB}{0, 0, 243}

\section*{Dataset Details}
In this section, we present the detailed description of the synthetic and real-world dataset used in this paper.
The synthetic datasets mainly refer to {\bfseries mLFR}\cite{mLFR}, which is a baseline model for generating multilayer networks. We also select ten common multilayer networks to conduct experiment. They are all collected manually and without ground-truth labels.

\begin{itemize}
\item {\texttt{\bfseries mLFR}}\cite{mLFR}: We mainly adjust the mixing parameter $\mu$ in mLFR datasets, which is the probability of edges connecting to the nodes of the other communities. A smaller $\mu$ indicates more obvious community structures, while a larger $\mu$ indicates more vague community structures. Table \ref{tab:mLFR} presents the detailed parameter settings of mLFR datasets in our experiment.
\end{itemize}

The ten real-world datasets were carefully chosen from domains like society, biology, finance, commerce, and others.

\begin{itemize}
\item {\texttt{\bfseries AUCs}}\cite{AUCs}: This multiplex social network contains five types of online and offline relationships (e.g., Facebook, Work, Lunch) among employees at a university's Computer Science department.
\item {\texttt{\bfseries C.Elegans}}\cite{CElegans}: This is the connectome of the Caenorhabditis elegans worm, where layers represent different types of synaptic junctions, such as electric and chemical connections.
\item {\texttt{\bfseries CKM}}\cite{CKM}: This dataset contains relationships among physicians in four towns in Illinois, originally collected to study how network ties influenced the adoption of a new drug.
\item {\texttt{\bfseries Drosophila}}\cite{Drosophila}\cite{SACCH}: This network represents different types of genetic interactions for organisms, sourced from the Biological General Repository for Interaction Datasets (BioGRID).
\item {\texttt{\bfseries EUAir}}\cite{EUair}: This is a transportation network composed of 37 layers, with each layer corresponding to a different airline operating in Europe.
\item {\texttt{\bfseries Kapferer}}\cite{kapferer}: This dataset documents interactions in a tailor shop in Zambia, where layers represent two different types of interactions recorded seven months apart.
\item {\texttt{\bfseries Lazega}}\cite{lazega2001collegial}\cite{lazega2006}: This multiplex social network consists of 3 kinds of interactions(Co-work, Friendship and Advice) between partners and associates of a corporate law partnership.
\item {\texttt{\bfseries RM}}\cite{RM}: This dataset records three social relationships (friendship and proximity at/outside work) between students at the MIT Media Laboratory and MIT Sloan business school.
\item {\texttt{\bfseries SACCHPOMB}}\cite{Drosophila}\cite{SACCH}: Sourced from the BioGRID repository, this network considers various types of genetic interactions for the fission yeast organism.
\item {\texttt{\bfseries Vickers}}\cite{vickers1981representing}: This social network data was collected from 29 seventh-grade students who were asked to nominate classmates based on several different relationships.
\end{itemize}

\section*{Baseline Details}
In this section, we present a detailed description of the selected nine state-of-the-art community detection methods, which can be divided into the following categories. The first group is traditional community detection algorithms:

\begin{itemize}
\item {\texttt{\bfseries GenLouvain(GL)}}\cite{GenLouvain}: A classic method that detects communities by maximizing modularity across layers, treating multilayer structure as a supra-adjacency matrix.
\item {\texttt{\bfseries M-Infomap(MI)}}\cite{MInfo}: A method based on the "Map Equation," which identifies communities by minimizing the description length of random walks.
\item {\texttt{\bfseries MDLPA}}\cite{MDLPA}: A multi-dimensional label propagation algorithm that extends the traditional method by propagating labels alternately between layers to jointly optimize the community partition.
\item {\texttt{\bfseries MolTi}}\cite{MolTi}: This algorithm identifies communities by measuring the consistency between network layers, then performs consensus clustering on the resulting similarity matrices.
\end{itemize}

The second group is five multiplex embedding algorithms.

\begin{itemize}
\item {\texttt{\bfseries HDMI}}\cite{HDMI}: A high-order deep learning method that captures complex structural features in multilayer networks and combines them with deep transfer learning for community detection.
\item {\texttt{\bfseries DMGI}}\cite{DMGI}: A contrastive learning-based algorithm that extends Deep Graph Infomax to multiplex networks to generate embeddings for community detection and other tasks
\item {\texttt{\bfseries GEAM}}\cite{GEAM}: A graph-enhanced attention model that uses an attention mechanism to integrate node attributes and structural information for generating high-quality node embeddings.
\item {\texttt{\bfseries NFACC}}\cite{NFACC}: A GCN-based method combines network feature augmentation with contrastive learning to integrate intra-layer and inter-layer information for robust multi-layer community detection.
\item {\texttt{\bfseries NACC}}\cite{NACC}: A self-supervised method fuses intra- and inter-layer features to create a learnable "anchor" network, which is then used in a contrastive learning framework to train the model.
\end{itemize}

\end{document}